%% file: main.tex
\documentclass[conference]{IEEEtran}
\IEEEoverridecommandlockouts
\usepackage{cite}
\usepackage{url}
\usepackage{amsmath,amssymb,amsfonts}
\usepackage{graphicx}
\usepackage{textcomp}
\usepackage{tabularx, booktabs}
\usepackage{multirow}
\usepackage[table,xcdraw]{xcolor}

\usepackage{listings}

\usepackage{adjustbox}
\usepackage{diagbox}
\usepackage{bm}

\usepackage[linesnumbered]{algorithm2e}
\SetAlCapSkip{1em}

\usepackage{tikz}
\usepackage{pgfplots}
\usetikzlibrary{pgfplots.groupplots,patterns}
\usetikzlibrary{shapes.geometric}
\usepackage[mode=buildnew]{standalone}

\tikzstyle{startstop} = [rectangle, rounded corners, minimum width=3cm, minimum height=1.25cm, text width=2.5cm, text centered, draw=black]

\tikzstyle{io} = [trapezium, trapezium left angle=70, trapezium right angle=110, minimum width=3cm, minimum height=1cm, text width=2.5cm, text centered, draw=black]

\tikzstyle{process} = [rectangle, minimum width=2.5cm, minimum height=1.25cm, text centered, text width=2.5cm, draw=black]

\tikzstyle{decision} = [diamond, minimum width=3cm, minimum height=1cm, text centered, draw=black]
\usetikzlibrary{arrows.meta}
\tikzstyle{arrow}= [thick, -{Stealth[length=3mm, width=2mm]}]

\def\BibTeX{{\rm B\kern-.05em{\sc i\kern-.025em b}\kern-.08em
    T\kern-.1667em\lower.7ex\hbox{E}\kern-.125emX}}
    
\newcolumntype{Y}{>{\centering\arraybackslash}X}

\usepackage[switch]{lineno}

\usepackage[absolute]{textpos}

\begin{document}

\title{
HEP-BNN: A Framework for Finding Low-Latency Execution Configurations of BNNs on Heterogeneous Multiprocessor Platforms
}


\author{

\IEEEauthorblockN{
Leonard David Bereholschi,
Ching-Chi Lin,
Mikail Yayla,
Jian-Jia Chen
}




 \IEEEauthorblockA{Technical University of Dortmund, Germany}
 \IEEEauthorblockA{\{leonard.bereholschi, chingchi.lin, mikail.yayla, jian-jia.chen\}@tu-dortmund.de}

}


\maketitle

\begin{textblock}{20}(1.2,0.3)
\noindent\small This paper has been accepted for presentation in the 5th Workshop on Accelerated Machine Learning (AccML), co-located with HiPEAC'23.
\end{textblock}

\thispagestyle{plain}
\pagestyle{plain}


\vspace{-10mm}
\begin{abstract}
Binarized Neural Networks (BNNs) significantly reduce the computation and memory demands 
with binarized weights and activations compared to full-precision NNs.
Executing a layer in a BNN on different devices of a heterogeneous multiprocessor platform consisting of CPU and GPU can affect the inference performance, i.e., accuracy and latency.
Usually, a heterogeneous HW platform consisting of a CPU and a GPU is available to execute the BNN workloads.
However, to use the heterogeneous HW effectively, it is necessary to find an efficient strategy for BNN workload mapping.
In this work, we propose a framework that generates efficient BNN layer-to-device mappings (i.e. suitable parallel configuration for each layer of the model) for execution platforms comprised of CPU and CUDA-capable GPU.
We evaluate our proposed framework with two BNN architectures using two well-known datasets, \emph{Fashion-MNIST} and \emph{CIFAR-10}, on three hardware platforms with different characteristics.
The results show that 
compared to running a fully-parallelized GPU implementation, our framework generates an efficient configuration up to $2\times$, $2.6\times$ and $11.8\times$ faster on our tested hardware respectively.
\end{abstract}

\begin{IEEEkeywords}
Binarized Neural Network, inference, GPU, CUDA
\end{IEEEkeywords}

\vspace{-5mm}
\section{Introduction}
\label{sec:introduction}

Neural Networks (NN) have been applied to various practical domains in the last decade, e.g., image recognition in computer vision, prediction of chemical patterns in chemistry, and cancer detection in medical science.~\cite{nnapp_abiodun2018}
Given a well-trained NN model, \textit{inference} is the process of using the model to make predictions against previously unseen data.
Depending on the structure of the NN model, the inference can be time- and resource-consuming. 

Binarized Neural Network (BNN)~\cite{Hubara/etal/2016} is a resource-efficient variant of NNs.
In BNNs, the weights and the activations are binarized into $1$-bit representation.
Multiplications and accumulations in a BNN can be computed using the \textit{xnor} operand and \textit{popcount()}, respectively.
Therefore, BNNs are significantly more resource-efficient compared to full-precision NNs, which makes BNNs excellent candidates for running AI application on resource-constrained edge devices.


For executing BNN workloads, customized hardware accelerators (i.e. on FPGAs or ASICs) are still in the early stages of development~\cite{nurvitadhi/etal/2016}, while \emph{CPUs and GPUs are mature and readily available}.
Several recent studies have evaluated the use of GPUs for BNNs.
Hubara et al.~\cite{Hubara/etal/2016} evaluate BNNs using \textit{XNOR} kernels for GPUs.
Xu et al.~\cite{xianda/etal/2019} implement a computation kernel for BNNs as well.
Li et al.~\cite{liang/etal/2021} run BNNs on Turing GPUs, focusing on the bit-level parallelism and strides in memory.
Chen et al.~\cite{chengang/etal/2020} develop a BNN acceleration engine for mobile phones.




Although GPU provides high computing capability, executing every layer in a BNN on GPU does not guarantee good performance.
We reveal in this work that executing all the BNN layers \textit{exclusively} on the GPU leads to significant increase in latency compared to running the model on the CPU sequentially (e.g. model is too small and CPU-overhead too significant, see Fig.~\ref{fig:cpu_vs_gpu}).
Therefore, a proper layer-to-device mapping is imperative for achieving efficient BNN inference on heterogeneous multiprocessor platforms consisting of CPU and GPU.

\textbf{Contributions}: In this paper, we propose a framework, \emph{HEP-BNN}, which automatically generates an efficient layer-to-device mapping (i.e. the suitable parallel configuration for each layer)
for a given BNN model running on a heterogeneous multiprocessor platform consisting of CPU and GPU.
Given a trained BNN model, \emph{HEP-BNN} systematically evaluates the execution time of the model on CPU and on GPU under different parallel configurations.
The configuration with the least execution time is highlighted as the efficient CPU/GPU configuration for the BNN on target platform.
Such a framework would lead to a more efficient use of available hardware resources.
Furthermore, automatically generating directly applicable code 
containing the optimized mapping, would enable highly efficient BNN inference.
It would allow both researchers and practitioners to fully exploit the capabilities of their available hardware platforms in applying BNNs efficiently on resource-limited edge devices.

Our contributions are summarized as follows:

\begin{itemize}
    \item 
    We present our \emph{HEP-BNN} framework that automatically generates the efficient layer-to-device mapping for given BNN models and heterogeneous execution platforms consisting of a CPU and a CUDA-capable GPU.
    The generated code, containing the efficient configuration for each layer of the model, can then be used for applications using BNN inference in practice.
    The proposed framework is published on Github\footnote{https://github.com/LeonardDavid/hep-bnn}.
    \item To demonstrate the capabilities of our framework, we apply our framework on two BNN models with two commonly used datasets, \textit{Fashion-MNIST} and \textit{CIFAR10}, on three hardware platforms with different characteristics: \textit{Server}, \textit{Laptop}, and \textit{Jetson TX2}. 
    Across the different BNN models, our results show that by applying properly chosen parallel configurations for layers running on GPU, inference times can be reduced by up to $2\times$, $2.6\times$ and $11.8\times$ for the three execution platforms, respectively.
\end{itemize}

\newcommand\titledistz{1.15}
\newcommand\figureWidthTz{9}
\newcommand\figureHeightTz{3.25}
\newcommand\barWidthTz{6}
\begin{figure}[t!]
    \begin{center}\scalebox{1}
    {
        \begin{tikzpicture}
            \definecolor{myblue}{RGB}{0, 112, 192}
            \definecolor{mygreen}{RGB}{0, 176, 80}
            \definecolor{myorange}{RGB}{255, 192, 0}
            \definecolor{myred}{RGB}{192, 0, 0}
            
            
             
             
             
            
            \begin{axis} [xbar = .05cm,
                bar width = 12pt,
                xmin = 0,
                xmax = 150,
                yticklabels ={,},
                ytick style={draw=none},
                height=\figureHeightTz cm,
                width=\figureWidthTz cm,
                bar width=1.7*\barWidthTz,
                xmajorgrids = true,
                xlabel = {Latency (s)},
                xlabel near ticks,
                xlabel shift = -0.15cm,
                xtick distance=25,
                scaled y ticks = false,
                ymin=-0.5,
                ymax=0.5,
                legend cell align=left,
                legend style={
                        at={(0.5,1.35)},
                        anchor=north,
                        column sep=1ex
                        },
                legend style={
                    nodes={scale=0.8, transform shape},
                    legend columns=-1,
                    },
                legend image post style={scale=0.5}
            ]
             
            \addplot[draw=myblue, fill=myblue] coordinates {(128,0) };
            \addplot[draw=myorange, fill=myorange] coordinates {(11.2,0) };
             
            \legend {GPU-only (parallel), CPU-only (sequential)};
             
            \end{axis}

        \end{tikzpicture}
        }
        \vspace{-0.25cm}
        \caption{
        {Fashion-MNIST on Jetson TX2: example from out results for the difference of total latency between the sequential CPU model and the parallel GPU model (with a higher CPU-overhead).
        }}
        \label{fig:cpu_vs_gpu}
        \vspace{-7.5mm}
    \end{center}
\end{figure}
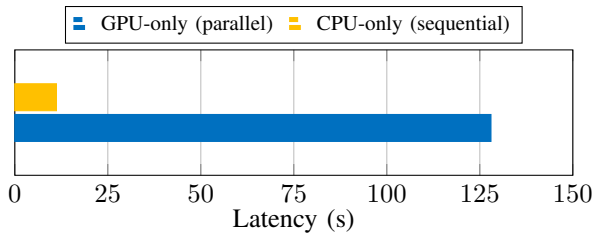

\begin{figure*}[tb]
    \centering
    \vspace{-10mm}
    \includegraphics[width=.95\textwidth]{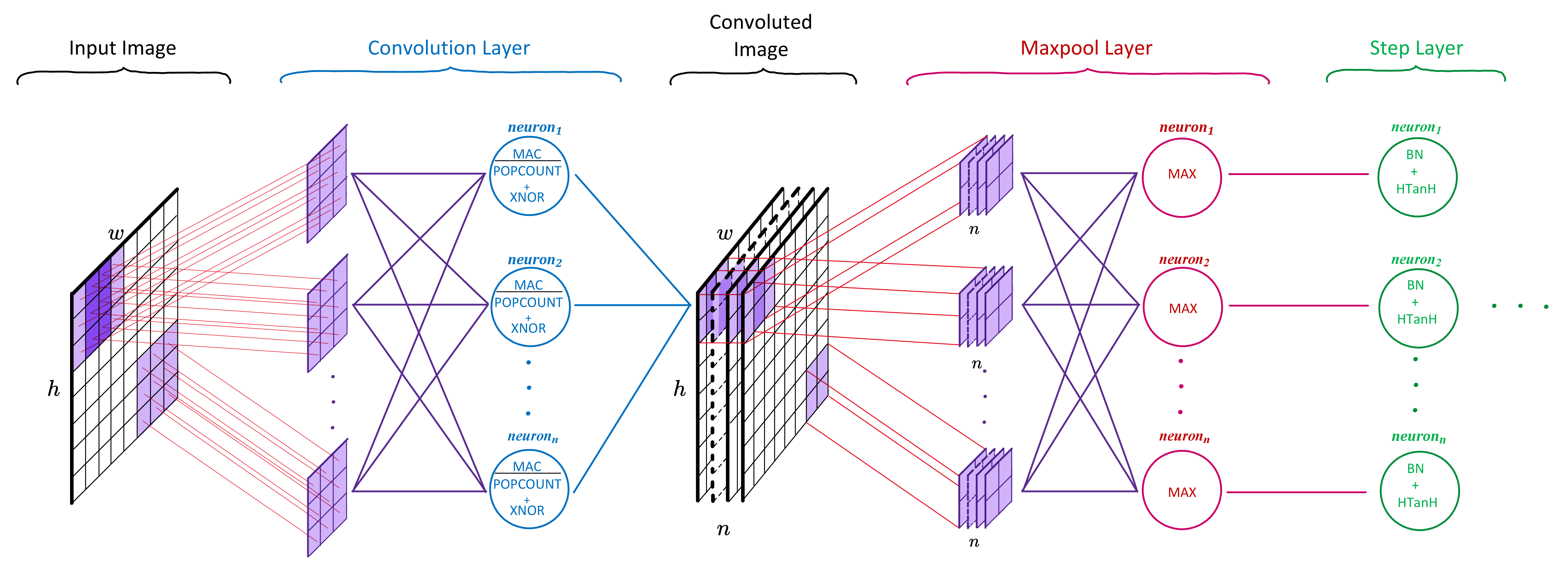}
    \vspace{-3mm}
    \caption{Structure of a Convolutional BNN model demonstrating the three major layer types: \textit{Convolution}, \textit{Maxpool}, and \textit{Step} layer.}
    \label{fig:nn_struct}
\end{figure*}

\section{System Model}


In Section~\ref{sec:bnn_basics}, the basics of BNNs the layers used in our models are presented.
An introduction into GPU computing is given in Section~\ref{sec:gpu}, where the CUDA framework is also described, including its features, limitations, and some implementation details.
The different parallel configurations implemented and used in our framework, as well as their notations used throughout this paper, are presented in Section~\ref{sec:parallelism_aspects}.
Finally, the problem definition is elaborated in Section~\ref{sec:problem_def}.

\subsection{Basics of Binarized Neural Networks}
\label{sec:bnn_basics}

Binarized Neural Networks (BNNs)~\cite{Hubara/etal/2016} are a resource-efficient variant of NNs.
In a BNN model, the weights and the activations are binarized into $1$-bit representation.
Unlike the full-precision NNs, where one matrix multiplication must be performed for computing the output of each neuron, we can simply apply $xnor$ operands for computing the outputs of neurons in BNNs.
Specifically, the output of a layer can be computed with $$2*{popcount}(xnor(W^l_i, I^{l-1})) - \#{bits} > T,$$ where $W^l_i$ are the weights of layer $l$, $I^{l-1}$ are the inputs to layer $l$, $popcount()$ is a function which counts the number of $1$s in the results of the $xnor$ operand, $\#bits$ is the number of bits in the $xnor$ operand, and $T$ is a learnable threshold parameter which can be computed with the batch normalization parameters.
The binary outputs depend on the truth value of the statement, which represents a shifted binarization function~\cite{sari/etal/2019}.


In this work, we consider convolutional BNNs.
There are four basic types of layers in a convolutional BNN, i.e., \textit{convolutional} layer, \textit{maxpool} layer, \textit{step} layer, and \textit{fully-connected} layer, as shown in Figure~\ref{fig:nn_struct}.
We also employ the \textit{flattening} layer in the BNNs in this work.

The \textit{convolutional} layer computes a 2D convolution of the input with filters.
We use ``C$\chi$'' to denote a \textit{convolutional} layer, where $\chi$ is
a number indicating the amount of neurons in the layer. 
For example, ``C$64$'' is a \textit{convolutional} layer with $64$ neurons.
In this work, the filter size is fixed at $3 \times 3$ for all the \textit{convolutional} layers.

The \textit{maxpool} layer downsamples the input by selecting the maximum value of the input in a given window size, which is set to $2 \times 2$ in our models.
We use ``MP$\chi$'' to denote a \textit{maxpool} layer, where $\chi$ indicates it's output size, e.g., ``MP16'' for a output of size ``$16\times16$''.

A \textit{step} layer performs batch normalization followed by a binary activation function.
Batch normalization~\cite{sari/etal/2019} is used in NN models for faster and more stable training, which helps increase the accuracy.
Note that the threshold values in the batch normalization are still signed integers, even in BNNs.
In our models, we apply Hard-Tanh as our binary activation function.
For inference in a BNN, batch normalization followed by activation can be computed with binary thresholding~\cite{sari/etal/2019}.
We denote the \textit{step} layer as ``S'' in the rest of this paper.

A \textit{fully-connected} layer connects every neuron in the current layer with every neuron in the next layer.
We denote a \textit{fully-connected} layer as ``FC$\chi$'', where $\chi$ is the amount of neurons in the layer.
Note that \textit{fully-connected} layers also have binary weights as learnable parameters.

We use ``FLAT'' to denote a \textit{flattening} layer.
The layer rearranges a high dimension matrix into a lower dimension matrix, e.g., from a $3$-dimension matrix into a $1$-dimension array in our models.
The rearrangement can be done with a simple one-line operation on CPU in C++ code.

\subsection{GPU}
\label{sec:gpu}

\begin{figure*}[t]
    \centering
    \vspace{-5mm}
    \includegraphics[width=.95\textwidth]{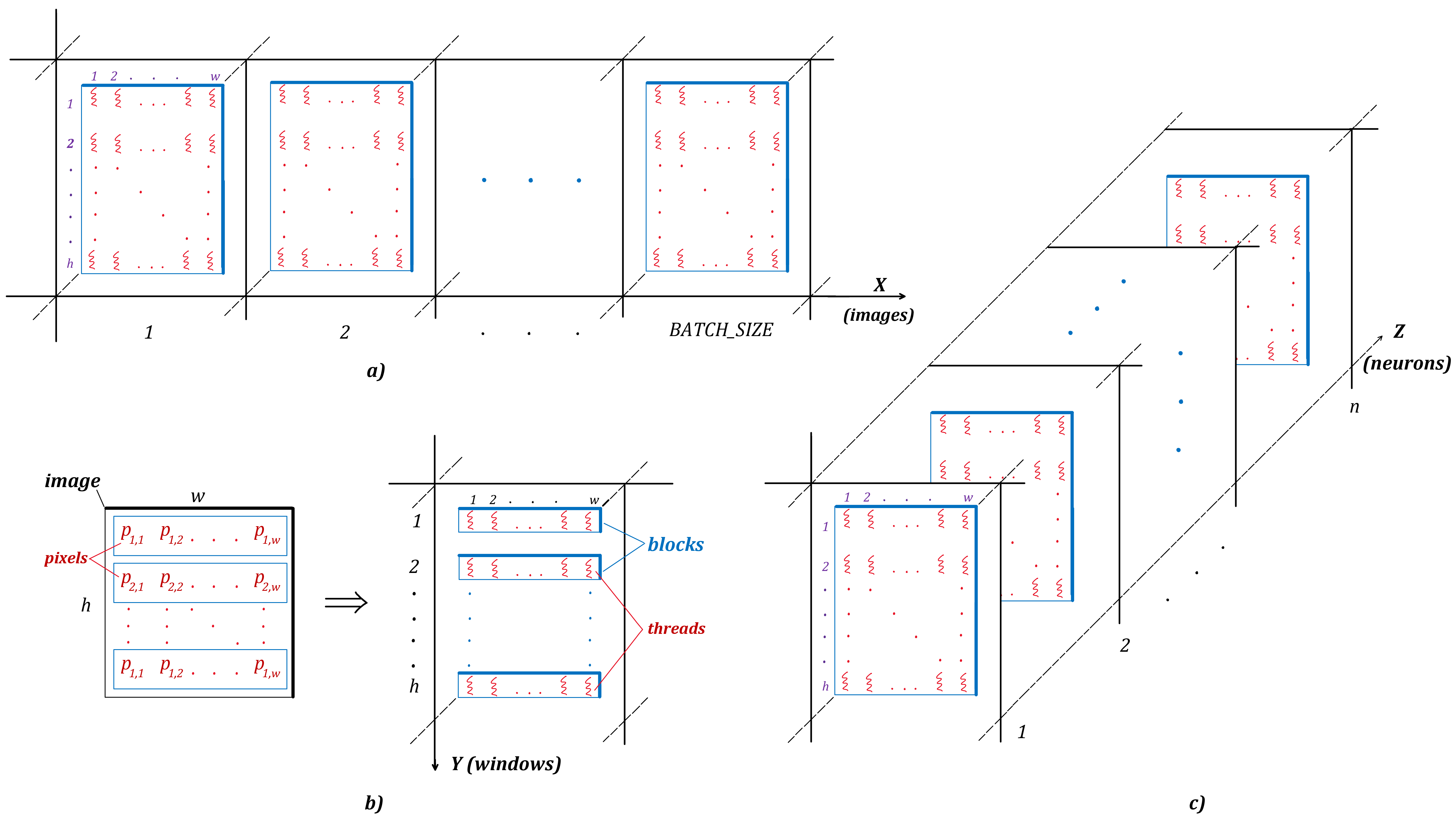}
    \vspace{-0.35cm}
    \caption{Concepts of the parallelism aspects: a) \emph{Data-based}, b) \emph{Window-based}, and c) \emph{Neuron-based}}
    \vspace{-5mm}
    \label{fig:separate_concepts_x_y_z}
\end{figure*}

Due to their architecture, GPUs are suitable for highly parallel use cases, such as repetitive matrix multiplication for graphics-intensive tasks. 
Most state-of-the-art GPUs have numerous computation cores, operating in an efficient manner with large and fast memories. 
On GPUs, the computations are performed by threads in parallel.
To program the thread behaviours, specialized GPU code, e.g., frameworks such as CUDA or OpenCL, needs to be employed.
In our work, we use the CUDA programming language to deploy the computation workload of a BNN model on Nvidia graphics cards. 

In a CUDA program, threads can be arranged into \textit{thread blocks}, in which up to $1024$ threads are executed in parallel.
Thread blocks can be further organized into a 3D \textit{grid} structure, allowing for a greater degree of parallelization on the GPU. 
The size of the grid is limited on the different dimension axes as follows: $\{x, y, z\} \rightarrow \{2^{31}-1, 65536, 65536\}$. 
Furthermore, the usage of internal CUDA variables, such as \textit{threadIDx} and \textit{blockIDx}, allow each thread to address individual values from arrays related to its specific computation.

Functions in which computation tasks are executed on GPU are called \emph{kernels} in the CUDA programming language.
Before launching a \emph{CUDA kernel}, memory is allocated on the GPU memory, after which all the required data for the computations is transferred from the host (CPU) to the device (GPU).
The sizes of the thread blocks and the grid are also specified before the \emph{kernel} launch, while respecting previously mentioned limitations.
After the GPU finishes executing every task in the \emph{kernel}, the results are copied from the device back to the host, and the previously allocated memory on the GPU is freed.

Although GPUs provide massive computational power compared to CPU, and are often used as accelerators in many use cases, running an application on the GPU does not always lead to performance improvement.
In some cases,
running parallel code on GPU can take longer than the sequential CPU code because of, for example, time overheads in communication.
Therefore, an analysis on the characteristics of an application, e.g., degree of parallelism, can determine if it is beneficial to run the application on GPU.
For applications that can be accelerated by GPU, how to organize the workload to achieve the optimal performance is a crucial issue that needs to be solved.

\subsection{Data Parallelism Aspects}
\label{sec:parallelism_aspects}

The workload of a BNN model consists of multiple data images which are used as input.
We define \emph{batch size} as the amount of data images in a batch, that are processed concurrently. 
To process the workload of BNN inference on a data set in parallel, we organize the workloads based on three aspects of data parallelism:
\begin{enumerate}
    \item \emph{Data-based}: every data image in a batch is inferred concurrently.
    \item \emph{Window-based}: a data image is divided into convolution windows of consecutive pixels, with the windows being processed concurrently. 
    \item \emph{Neuron-based}: the outputs of the neurons in the same layer in a NN model are calculated concurrently.
\end{enumerate}

In the \emph{Data (X)} configuration, multiple data images in a batch are inferred on the GPU concurrently, as shown in Figure~\ref{fig:separate_concepts_x_y_z} (a). 
Each data image in a batch is assigned to one GPU thread block.
If a thread block is assigned with multiple data images, these images are processed one after another.
Each pixel (and its subsequent operations) in a data image is processed by one thread in the thread block.


Figure~\ref{fig:separate_concepts_x_y_z} (b) demonstrates the idea for the \emph{Window (Y)} configuration, in which a data image is divided into windows of consecutive pixels in a row-wise manner, with each window being assigned to one GPU thread block.
The workloads related to the pixels (threads) in the same window are processed on the GPU concurrently.


For the \emph{Neuron (Z)} configuration, the outputs of neurons from the same layer are calculated concurrently, as shown in Figure~\ref{fig:separate_concepts_x_y_z} (c).
The output of a neuron is the weighted sum from its predecessors after going through an activation function.
Each neuron in a layer is assigned to a GPU thread block, with threads in a thread block taking the corresponding outputs from the previous layer as input, and calculating the output for the neuron.
Using these aspects, we consider the following seven parallel configurations and their notations, which will be used throughout this paper: 1) \emph{Data (X)}, 2) \emph{Window (Y)}, 3) \emph{Neuron (Z)}, 4) \emph{Data + Window (XY)}, 5) \emph{Data + Neuron (XZ)}, 6) \emph{Window + Neuron (YZ)}, and 7) \emph{Data + Window + Neuron (XYZ)}.
The configurations composed of multiple aspects are implemented according to all of the implementations of the individual aspects at the same time.
Note that for all the parallel configurations, the threads in thread blocks perform the same operation depending on the layer, e.g., convolution of pixels in a convolution layer.
The \textit{blockIDx} and \textit{threadIDx} variables determine which pixel(s) and/or neuron(s) each thread is responsible for.

\vspace{-2mm}
\subsection{Problem Definition}
\label{sec:problem_def}


Given a well-trained BNN model, we aim to reduce the inference time of a BNN model with the help of GPU.
However, there are multiple aspects for parallelizing the computation workloads on GPU.
Each layer in the BNN model can have different suitable parallel configurations. 
Nevertheless, there might also be layers with workloads that are not beneficial if running on GPU due to overheads, e.g., which are caused by data migration between host and the GPU device.
Therefore, our objective is to generate an efficient layer-to-device mapping for a given BNN model, so that the inference time is minimized.
For layers that are mapped to GPU for execution, we also determine their suitable parallel configurations.  


\vspace{-1mm}
\section{Framework Presentation}
\label{sec:framework_presentation}

We introduce the proposed framework in details in this section.
The operational steps of our \emph{HEP-BNN} framework are outlined in Section~\ref{sec:our_framework}.
The mapping algorithm is described in Section~\ref{sec:algorithm}.
Information about the folder structure, important script files, and generated files of the framework, are detailed in Sections~\ref{sec:folder_structure},~\ref{sec:script_files} and~\ref{sec:generated_files}, respectively.

\vspace{-2mm}
\subsection{High-level Overview of Our \emph{HEP-BNN} Framework}
\label{sec:our_framework}

\begin{figure}[t]
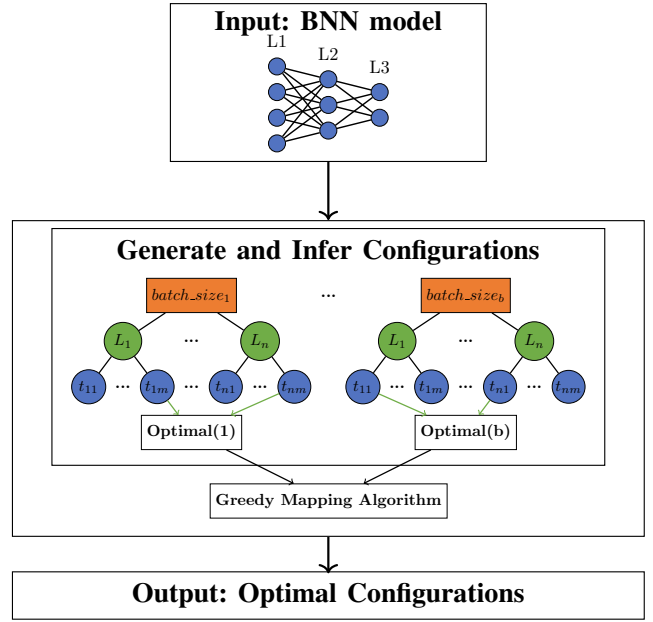

    \centering 
    \includestandalone[width=.95\columnwidth]{figures/fi_op_diagram}
    \vspace{-2mm}
    \caption{Operational steps our \emph{HEP-BNN} framework.}
    \vspace{-3mm}
    \label{fig:operational_diagram}
\end{figure}

The operational steps performed by our framework are represented in Figure~\ref{fig:operational_diagram}.
First, the program receives a BNN model in ONNX format as input, previously trained on a specific dataset (e.g. Fashion-MNIST, CIFAR10). 
Then, for every batch size in a defined range, the appropriate C++ and CUDA code for the CPU and GPU is generated.
After every layer ($L_1, \ldots, L_n$) is implemented using different configurations,
the model is inferred for every configuration applied, which results in different runtimes (with~$t_{11}, \ldots, t_{1m} \in L_1~\text{and}~t_{n1}, \ldots, t_{nm} \in L_n$).
These timing information are used for choosing an efficient configuration 
in a greedy manner, according to Alg.~\ref{alg:framework} described in Section~\ref{sec:algorithm}.



\vspace{-2mm}
\subsection{Mapping Algorithm}
\label{sec:algorithm}




In order to determine the suitable \emph{parallel configuration} which achieves the lowest inference time, the following layer-to-device mapping algorithm is applied (see Alg.~\ref{alg:framework}).
The algorithm profiles each layer of the BNN model using different batch sizes, both on the CPU and the GPU.
On the GPU, every parallel configuration from Section~\ref{sec:parallelism_aspects} are implemented.
In total, each layer is profiled on 8 different implementations: 1) \emph{CPU}, 2) \emph{\mbox{Data (X)}}, 3) \emph{\mbox{Window (Y)}}, 4) \emph{\mbox{Neuron (Z)}}, 5) \emph{\mbox{Data + Window (XY)}}, 6) \emph{\mbox{Data + Neuron (XZ)}}, 7) \emph{\mbox{Window + Neuron (YZ)}}, and 8) \emph{\mbox{\mbox{Data + Window} + Neuron (XYZ)}}. 


The \emph{implementation} that achieves the lowest inference time for the profiled \emph{layer} is mapped to the specific \emph{batch size}.
Summing up the lowest inference times for every \emph{layer}, results in the total runtime for executing the BNN model, using the \emph{efficient implementations} for the \emph{specific batch size}.

After profiling every \emph{layer}, the \textbf{minimal} total runtime, as well as the \emph{proper batch size} for which it is achieved, is searched.
Finally, the \emph{proper batch size} is used to get the mapped \emph{implementation} of each \emph{layer} of the BNN model.
This creates the \emph{efficient parallel configuration}, which achieves the lowest expected inference time.
Algorithm~\ref{alg:framework} represents the pseudocode of the described mapping algorithm.
\begin{algorithm}[h]
\KwData{BNN model}
\KwResult{Efficient Configuration}
$proper_{batch\_size} \gets 0$

$result\_time \gets \texttt{MAX\_INT}$

\ForEach{$batch\_size$}{
    $sum_{min\_time} \gets 0$
    
    \ForEach{$layer$}{
        $min\_time \gets \texttt{MAX\_INT}$
        
        \ForEach{$implem$}{
            implement $layer$ using $implem$
            
            (CPU$_{time}$, GPU$_{time}$) $\gets$ profile(implemented$_{layer}(batch\_size)$)

            $inference\_time \gets$ CPU$_{time}$ + GPU$_{time}$
            
            \If{$inference\_time < min\_time$}{
                $min\_time \gets inference\_time$
                
                MAP $implem(layer)$ to $batch\_size$
            }
        }
        $sum_{min\_time} \gets sum_{min\_time} + min\_time$
    }
    \If{$sum_{min\_time} < result\_time$}{
        $result\_time \gets sum_{min\_time}$
        
        $proper_{batch\_size} \gets batch\_size$
    }
}

\ForEach{$layer$}{
    get $implem(layer)$ from MAP[$proper_{batch\_size}$]
    
    add $layer_{implem}$ to \textit{Efficient Configuration}
}

\KwRet{\textit{Efficient Configuration}}

\caption{Mapping algorithm that determines the efficient configuration of a BNN model that achieves the lowest inference time (Note: \textit{implem} is short for \textit{implementation})
}
\vspace{-2mm}
\label{alg:framework}
\end{algorithm}

\subsection{Folder Structure}
\label{sec:folder_structure}

Our \emph{HEP-BNN} framework uses Python to run the implementations and optimizations of the input model.
To exemplify our implementation, we use the open source machine learning compiler and code generator Fastinference
~\cite{fastinference}.
Specifically, for the generation of the C++ and CUDA code for the CPU and GPU respectively, the templating language \textit{Jinja2} is used.
An overview of the most important part of the folder-tree structure (\textit{'fastinference/'}) will be outlined in this section. 

Each model, optimizer and implementation is defined in a separate folder in \textit{'fastinference/'}. 
These are further separated into the supported algorithm types such as \textit{'ensemble/'}, \textit{'tree/'}, \textit{'neuralnet/'}. 
Specifically, in \textit{'fastinference/implementations/neuralnet'} there are folders for the different target hardware: \textit{'cpp/'}, \textit{'fpga/'}, \textit{'iree/'}. 

This is where the main part of our work is located, namely in the \textit{'cuda/'} folder, which contains a separate directory for each parallel configuration. The folder names follow the notations introduced 
in Section~\ref{sec:parallelism_aspects}.

In every folder there are \textit{Jinja2} files containing mainly CUDA code templates for every layer, for parallel execution of the model on the GPU.
There is also a \textit{'cpu/'} folder, that contains C++ templates for the sequential operation of the model on the CPU, used for the sequential implementations.
Each \textit{'implement.py'} file found in every folder, contains the function \textit{'to\_implementation()'} and is responsible for selecting the appropriate template files for each layer of the model, and to generate the necessary C++ and CUDA files for the implementation.

Additionally, the \textit{'automatic/'} folder contains the code which runs our mapping algorithm, that automatically reads the model and data from the appropriate path, generates and infers all of the selected configurations, and maps the suitable configuration in a greedy manner.
In the following section, we give a more detailed description of the usage of certain files, by using the CIFAR10 dataset as an example.

\vspace{-2mm}
\subsection{Important script files}
\label{sec:script_files}

In \textit{\textbf{test\_cuda.py}}, the implementations of the parallel configurations which can be used, are specified. Their notations are consistent with the ones described in Section~\ref{sec:parallelism_aspects}. 
Afterwards, the \emph{HEP-BNN} framework is launched by calling the \textit{'to\_implementation()'} function found in \textit{\textbf{test\_utils.py}}. 

Here, an upper and lower bound is set for the batch sizes, which are expressed as powers of 2 (e.g. with 'b\_l = 0' and 'b\_u = 4', the batch sizes used are $2^{0}=1$, $2^{1}=2$, $2^{2}=4$, and $2^{3}=8$).

The mapping algorithm is then run inside of the nested for-loops, one for each configuration, and the other one for each batch size.
It consists of two important functions, \textit{'prepare\_fastinference()'} and \textit{'run\_experiment()'}.
The former generates all the necessary files for the model to compile and run (more details will be given in section~\ref{sec:generated_files}), while the latter function compiles and runs the generated model, after which it outputs and stores the results of the inference.

After running all of the generated models, the best configuration for each layer is mapped according to the lowest inference time (see Alg.~\ref{alg:framework} -- lines 23:26).
Finally, the efficient configuration is generated and inferred, achieving the lowest inference time out of all other combinations.



    
    
    






The BNN model is stored in ONNX format under the following path:
\vspace{-1mm}
\begin{verbatim}
fastinference/implementations/
neuralnet/cuda/automatic/model/
cifar10/model_cifar10.onnx
\end{verbatim}
\vspace{-2mm}



Test data is stored under:
\vspace{-1mm}
\begin{verbatim}
fastinference/implementations/
neuralnet/cuda/automatic/data/
cifar10/testing.csv
\end{verbatim}
\vspace{-2mm}



\emph{HEP-BNN} is \textbf{launched} by running the following command in the root folder (\textit{'fastinference/'}): 
\vspace{-1mm}
\begin{verbatim}
fastinference/implementations/neuralnet/
cuda/automatic/test_cuda.py 
--outpath tmp/fastinference/cuda_auto 
--dataset cifar
\end{verbatim}



\vspace{-2mm}
\subsection{Generated files}
\label{sec:generated_files}
In this section, we present the list of \textbf{generated files} for each configuration, including a brief description.

\begin{itemize}
    \item \textit{'utils.h'} and \textit{'utils.cuh'}: contain utility functions (for C++ and CUDA code respectively)
    \item \textit{'cuda\_kernel.h'} and \textit{'cuda\_model.h'}: are headers containing functions that link the C++ code to the CUDA code
    \item \textit{'modelW.hpp'}: contains declarations of the output arrays for every layer, and stores the weights, biases, and thresholds
    \item \textit{'model.h'} and \textit{'model.cpp'}: depending on the configuration, contains either the sequential C++ model for the CPU, or the calls to parallel CUDA model for the GPU (both implementations are present, but the unused part is commented out for debugging and comparison purposes)
    \item \textit{'model.cu'}: CUDA code for the GPU (if applicable)
\end{itemize}






There are also a few files which are the same, regardless of configuration, which are already written and \textbf{copied} from the \textit{'../cuda/automatic/'} folder to every generated implementation:

\begin{itemize}
    \item \textit{'main.cpp'}: splits the dataset into batches, calls the model predictor and calculates the accuracy and latency
    \item \textit{'CMakeLists.txt'}: generates the 'Makefile' that compiles the entire project
\end{itemize}




Note that file changes to any of the \textit{Jinja2} templates and 'implement.py' files require the following command to be run \textit{'python setup.py install'}, before calling the \textit{'make'} command.

\begin{table}[t]
  \centering
  \caption{Structure of the CIFAR-10 BNN Models. 
  }
  \vspace{-2mm}
  \label{tab:nn_struct_cifar}
  \begin{tabularx}{\linewidth}{cc}
  \toprule
  & CIFAR-10 BNN model Structure  \\ \midrule
  & In $\to$ C64 $\to$ S $\to$ C64 $\to$ MP16 $\to$ S $\to$ C256 $\to$ S $\to$ C256\\
  & \phantom{In} $\to$ MP8 $\to$ S $\to$ C512 $\to$ S $\to$ C512 $\to$ MP4 $\to$ S \\
  & \phantom{In} $\to$ FLAT $\to$ FC1024 $\to$ S $\to$ FC1024 $\to$ 10\\ \bottomrule
  \vspace{-7.5mm}
  \end{tabularx}
\end{table}

\begin{table}[t]
  \centering
  \caption{
  Structure of the FashionMNIST BNN Models. 
  }
  \vspace{-2mm}
  \label{tab:nn_struct_fashion}
  \begin{tabularx}{\linewidth}{ccc}
  \toprule
  \textcolor{white}{LDB} & & FashionMNIST BNN model structure  \\ \midrule
  \phantom{707} & & In $\to$ C64 $\to$ MP14 $\to$ S $\to$ C64 $\to$ MP7 $\to$ S \\
  & & \phantom{In} $\to$ FLAT $\to$ FC2048 $\to$ S $\to$ FC2048 $\to$ 10 \\ \bottomrule
  \vspace{-7.5mm}
  \end{tabularx}
\end{table}

\begin{table}[t]
    \centering
    \caption{Overview of Hardware used for evaluation}
    \label{tab:hw_overview}
    \vspace{-2mm}
    \begin{tabularx}{\linewidth}{{ll|YYY}}
    \toprule
        & Name & CPU & GPU & CUDA Cores \\ \midrule
        & Server & i7-8700K & GTX1080 & 2560\\ 
        & Laptop & i7-10750H & GTX1650Ti & 1024\\ 
        & Jetson TX2 & Cortex-A57 & Pascal-based & 256 \\ 
    \bottomrule
    
    \vspace{-9mm}
    \end{tabularx}
\end{table}

\vspace{-1mm}
\section{Experiment Setups and Results / Evaluation}\label{sec:eval}




In this section, we present our experimental results.
Information about the hardware, datasets, and models are presented in Section~\ref{sec:prof_env}.
In Section~\ref{sec:results}, the results of our \emph{HEP-BNN} framework, i.e., implementing the suitable configurations for every layer, are presented and compared to the baseline sequential implementation, as well as other parallel configurations.


\begin{table*}[t]
\centering
\caption{Efficient configurations for the CIFAR10 model}
\vspace{-2mm}
\label{tab:optimal_cifar}
\resizebox{\textwidth}{!}{%
\begin{tabular}{lccccccccccccccccccc}
\hline
 &
  {\color[HTML]{333333} \textbf{C64}} &
  {\color[HTML]{333333} S} &
  {\color[HTML]{333333} \textbf{C64}} &
  \textbf{MP16} &
  S &
  \textbf{C256} &
  S &
  \textbf{C256} &
  \textbf{MP8} &
  S &
  \textbf{C512} &
  S &
  \textbf{C512} &
  \textbf{MP4} &
  S &
  FLAT &
  \textbf{FC1024} &
  S &
  \textbf{FC1024} \\ \hline
\rowcolor[HTML]{E0E0E0} 
Server &
  {\color[HTML]{333333} \textbf{CPU}} &
  {\color[HTML]{333333} CPU} &
  {\color[HTML]{333333} \textbf{Z}} &
  \textbf{CPU} &
  CPU &
  \textbf{XZ} &
  CPU &
  \textbf{XYZ} &
  \textbf{XY} &
  CPU &
  \textbf{XZ} &
  CPU &
  \textbf{XZ} &
  \textbf{X} &
  CPU &
  CPU &
  \textbf{X} &
  CPU &
  \textbf{CPU} \\
Laptop &
  {\color[HTML]{333333} \textbf{CPU}} &
  {\color[HTML]{333333} CPU} &
  {\color[HTML]{333333} \textbf{Y}} &
  \textbf{CPU} &
  CPU &
  \textbf{XYZ} &
  CPU &
  \textbf{XYZ} &
  \textbf{CPU} &
  CPU &
  \textbf{XYZ} &
  CPU &
  \textbf{XYZ} &
  \textbf{CPU} &
  CPU &
  CPU &
  \textbf{X} &
  CPU &
  \textbf{CPU} \\
\rowcolor[HTML]{E0E0E0} 
TX2 &
  {\color[HTML]{333333} \textbf{CPU}} &
  {\color[HTML]{333333} CPU} &
  {\color[HTML]{333333} \textbf{XYZ}} &
  \textbf{CPU} &
  CPU &
  \textbf{Z} &
  CPU &
  \textbf{Z} &
  \textbf{CPU} &
  CPU &
  \textbf{XZ} &
  CPU &
  \textbf{XZ} &
  \textbf{CPU} &
  CPU &
  CPU &
  \textbf{XY} &
  CPU &
  \textbf{CPU} \\ \hline
  \vspace{-7.5mm}
\end{tabular}%
}
\end{table*}

\begin{table}[t]
\centering
\vspace{-2mm}
\caption{Efficient configurations for Fashion-MNIST model}
\vspace{-2mm}
\label{tab:optimal_fashion}
\resizebox{\columnwidth}{!}{%
\begin{tabular}{lcccccccccc}
\hline
 &
  {\color[HTML]{333333} \textbf{C64}} &
  {\color[HTML]{333333} \textbf{MP14}} &
  {\color[HTML]{333333} S} &
  \textbf{C64} &
  \textbf{MP7} &
  S &
  FLAT &
  \textbf{FC2048} &
  S &
  \textbf{FC2048} \\ \hline
\rowcolor[HTML]{E0E0E0} 
Server &
  {\color[HTML]{333333} \textbf{CPU}} &
  {\color[HTML]{333333} \textbf{CPU}} &
  {\color[HTML]{333333} CPU} &
  \textbf{XZ} &
  \textbf{X} &
  CPU &
  CPU &
  \textbf{CPU} &
  CPU &
  \textbf{CPU} \\
Laptop &
  {\color[HTML]{333333} \textbf{CPU}} &
  {\color[HTML]{333333} \textbf{CPU}} &
  {\color[HTML]{333333} CPU} &
  \textbf{CPU} &
  \textbf{CPU} &
  CPU &
  CPU &
  \textbf{CPU} &
  CPU &
  \textbf{CPU} \\
\rowcolor[HTML]{E0E0E0} 
TX2 &
  {\color[HTML]{333333} \textbf{CPU}} &
  {\color[HTML]{333333} \textbf{CPU}} &
  {\color[HTML]{333333} CPU} &
  \textbf{XZ} &
  \textbf{CPU} &
  CPU &
  CPU &
  \textbf{CPU} &
  CPU &
  \textbf{CPU} \\ \hline
  \vspace{-10mm}
\end{tabular}%
}
\end{table}

\vspace{-2mm}
\subsection{Profiling Environment}
\label{sec:prof_env}

\subsubsection{\textbf{Hardware}}
We execute our experiments on a Linux based server equipped with an Intel Core i7-8700K CPU and a GTX1080 GPU with 8GB of video memory each.
Additionally, we run the same experiments on a Windows-system with a consumer-grade GPU (GTX 1650Ti), as well as on an embedded Jetson TX2 board.
Table~\ref{tab:hw_overview} also lists the amount of available CUDA cores for each GPU.

Kernel launches are wrapped around \texttt{cudaEventRecord()} functions, in order to accurately measure the GPU-time.
The communication cost (i.e. memory allocation and transfer between host and device) in the CUDA code is executed before the kernel launch, by the CPU, and is included in the CPU-overhead time.
We implement independent layers for the models, therefore data transfer between CPU and GPU takes place before and after every layer’s execution (even if two consecutive layers are executed on the GPU). 
The code generator can be adapted to consider this case in future works/implementations. 


\subsubsection{\textbf{Datasets}}
The algorithm is evaluated on two commonly used benchmarking datasets, namely \mbox{FashionMNIST} and \mbox{CIFAR-10}. 
The \mbox{FashionMNIST}~\cite{Fashion-MNIST} dataset consists of $70,000$ gray-scale images and labels from $10$ classes, representing different clothing articles. 
The size of each image is $28\times28$ pixels in $1$ channel, with $0$ representing the \emph{brightest} and $255$ the \emph{darkest} values.
Out of the total amount of images, $60,000$ are used for training, while the remaining $10,000$ for testing.
The \mbox{CIFAR-10}~\cite{cifar10} dataset contains $60,000$ colour images ($3$ channels), each with a size of $32\times32$ for a total of $1024$ pixels. 
It is split into $50,000$ training and $10,000$ test images, which are classified in $10$ different classes representing means of transportation (i.e. airplane, ship, truck, automobile) and animals (i.e. bird, cat, dog, deer, frog, horse).

\subsubsection{\textbf{BNN Architecture}}
The BNN models used for inferring the datasets are VGG-type architectures~\cite{simonyan/etal/2014} adapted for the binarized variant of NNs.
The \mbox{FashionMNIST} network model contains a total of $10$ layers, each of them belonging to one of the types presented in Section~\ref{sec:bnn_basics}.
Specifically, the $1^{st}$ and $4^{th}$ layer are convolutional layers, with a size of $28\times28\times64$ and $14\times14\times64$ respectively.
The convolutional layers are down-sampled to half of their input size, in the immediately following maxpool layers, namely in the $2^{nd}$ and $5^{th}$ layer.
Step layers are employed on the $3^{rd}$, $6^{th}$, and $9^{th}$ layer, which apply batch normalization and the activation function.
After convoluting and down-sampling the input image, it is flattened into a 1-dimensional array in layer $7$.
Finally, a total of $2048$ neurons are fully-connected in the $8^{th}$ and $10^{th}$ layer.




The structures of the networks are listed in Table~\ref{tab:nn_struct_fashion} and~\ref{tab:nn_struct_cifar}, using the notations from Section~\ref{sec:bnn_basics}.
The \mbox{CIFAR-10} model also contains the standard layer types for BNNs, totalling to $19$ layers.
Convolutional layers are placed on the $1^{st}$, $3^{rd}$, $6^{th}$, $8^{th}$, $11^{th}$, and $13^{th}$ position.
Down-sampling using maxpooling occurs only three times, namely in the $4^{th}$, $9^{th}$, and $14^{th}$ layer.
The $16^{th}$ layer flattens the image for the fully-connected layers at position $17$ and $19$.
The rest of the layers are step layers.

To simulate the inference process, the sets of test images from both models respectively are used.
This guarantees a controlled benchmarking environment for the algorithm that profiles each layer of the BNN models.
The weights and biases were trained over the course of $100$ epochs, and achieve an inference accuracy of $77.24\%$ for the \mbox{FashionMNIST}, and $67.08\%$ for the \mbox{CIFAR-10} model.

Preliminary observations showed that the batch size has an impact on runtime for certain configurations.
Therefore, the experiments also apply different batch sizes for the two BNN models, from $\{1, 2, ..., 128\}$, in increments of powers of 2.

\vspace{-2mm}
\subsection{Results}
\label{sec:results}

Table~\ref{tab:results} presents the inference times measured by running the BNN models (each with $10000$ data images as input), for every tested target hardware. 
The latency (displayed in seconds) is the time required by the BNN to process the \textbf{entire} test dataset of $10000$ images.
Recall from Section~\ref{sec:algorithm}, that the suitable configuration is chosen over all the different batch sizes, and therefore it is important to note the batch size alongside the runtime.
Specifically, a batch size of $n$ means that $n$ data-images are processed in parallel. 

It can be observed that the server, which has the most CUDA cores out of the three tested hardware, has overall faster runtimes.
On the other hand, the resource-constrained TX2, having the least amount of CUDA cores, has significantly higher runtimes.





The suitable configurations mapped for each layer is presented in Tables~\ref{tab:optimal_cifar} and~\ref{tab:optimal_fashion} for the \emph{\mbox{CIFAR-10}} and \emph{\mbox{FashionMNIST}} models respectively.
From there, we can notice the following observations:
\begin{itemize}
    \item Since the Fashion-MNIST BNN model is smaller out of the two, almost all of the layers are mapped to the CPU for sequential execution. A notable exception is the second convolution layer, which is mapped to the \textit{XZ} configuration on the Server and the TX2.
    \item In the case of the CIFAR10 BNN model, we observe that the maxpool layer is mapped to the CPU in almost every case, whereas the convolutional and fully-connected layers are mapped to the GPU using different parallel configurations. For example, the second convolutional layer is mapped to the \textit{Z} configuration on the server, \textit{Y} on the Laptop and \textit{XYZ} on the TX2.
\end{itemize}

\begin{figure*}[t!]
    \vspace{-5mm}
    \begin{center}\scalebox{1}{
        \begin{tikzpicture}
            \definecolor{myblue}{RGB}{0, 112, 192}
            \definecolor{mygreen}{RGB}{0, 176, 80}
            \definecolor{myorange}{RGB}{255, 192, 0}
            \definecolor{myred}{RGB}{192, 0, 0}
            \begin{groupplot}[group style = {group size = 1 by 6, horizontal sep =15pt, vertical sep = 30 pt}]
            \input{figures/tikz/server_cifar}
            \vspace{-5mm}
            \input{figures/tikz/server_home_cifar}
            \input{figures/tikz/server_tx2_cifar}
            
            \input{figures/tikz/server_fashion}
            \input{figures/tikz/server_home_fashion}
            \input{figures/tikz/server_tx2_fashion}
            \end{groupplot}    
        \end{tikzpicture}}
        \caption{
        {Execution time over batch size comparison for the entire FashionMNIST test images dataset (upper three figures) and entire CIFAR10 test images dataset (lower three figures). Each dataset is evalauted with three different hardware configurations, namely: Server, Laptop, and TX2.}}
        \label{fig:graphs_final}
        \vspace{-7mm}
    \end{center}
\end{figure*}
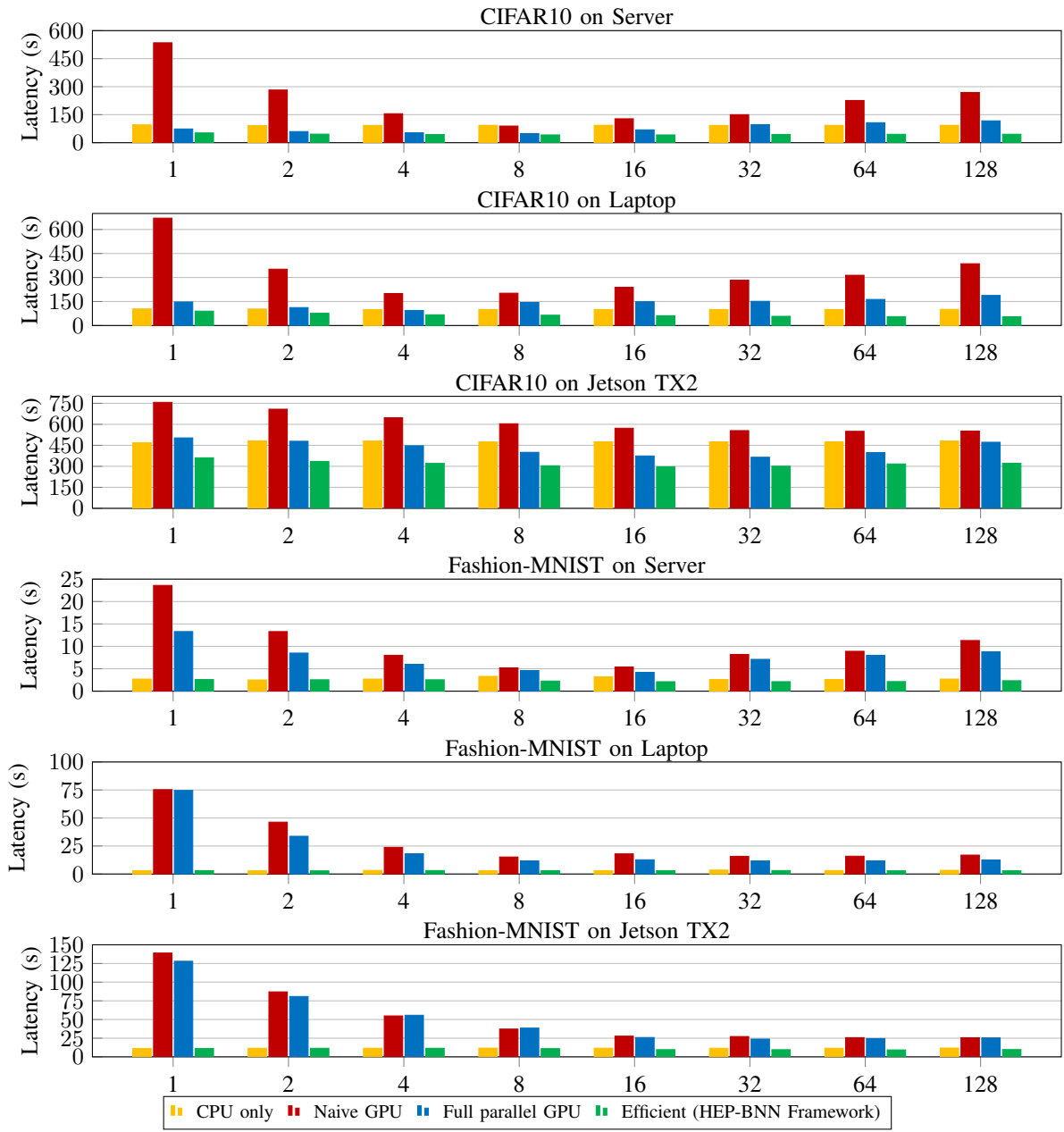

Finally, Figure~\ref{fig:graphs_final} compares the purely sequential CPU implementation and two intuitive ways of parallelizing the BNN models to the results of the mapping algorithm.
The naive GPU implementation considers the parallelization of every layer suitable for GPU acceleration using only the \emph{\mbox{Data (X)}} configuration.
The full-parallel GPU implementation parallelizes every suitable layer as much as possible, i.e., applying the \emph{\mbox{\mbox{Data + Window} + Neuron (XYZ)}} configuration.
These are the two most intuitive ways of parallelizing the workload, while not considering the fact that some layers may not benefit from GPU acceleration.

The results from running the efficient configurations for every batch size, show a significant speedup overall.
Specifically, compared to the fully-parallel implementation, running the \emph{HEP-BNN} framework on the server leads to at most $2\times$ speedup, while on the Jetson TX2, it achieves at most $2.6\times$ speedup, and on the Laptop-system the efficient configuration results in a $11.8\times$ improvement.

\begin{table}[t]
\centering
\vspace{-2mm}
\caption{The minimum inference times of the efficient configurations}
\vspace{-2mm}
\label{tab:results}
\resizebox{\columnwidth}{!}{%
\begin{tabular}{lcccccc}
\hline
\rowcolor[HTML]{FFFFFF} 
\multicolumn{1}{c|}{\cellcolor[HTML]{FFFFFF}Dataset} & \multicolumn{3}{c|}{\cellcolor[HTML]{FFFFFF}Fashion-MNIST} & \multicolumn{3}{c}{\cellcolor[HTML]{FFFFFF}CIFAR10} \\
\rowcolor[HTML]{E0E0E0} 
\multicolumn{1}{c|}{\cellcolor[HTML]{E0E0E0}Hardware} & Server & Laptop & \multicolumn{1}{c|}{\cellcolor[HTML]{E0E0E0}TX2}   & Server & Laptop & TX2  \\ \hline
\rowcolor[HTML]{FFFFFF} 
\multicolumn{1}{l|}{\cellcolor[HTML]{FFFFFF}Runtime}  & 2.11s  & 2.84s  & \multicolumn{1}{c|}{\cellcolor[HTML]{FFFFFF}9.31s} & 41.6s  & 55s    & 297s \\
\rowcolor[HTML]{E0E0E0} 
\multicolumn{1}{l|}{\cellcolor[HTML]{E0E0E0}Batch size}  & 16  & 2  & \multicolumn{1}{c|}{\cellcolor[HTML]{E0E0E0}64} & 16  & 128 & 16 \\
\hline
\end{tabular}
}
\vspace{-7mm}
\end{table}

\newcommand\titledist{1.15}
\newcommand\figureWidthT{16}
\newcommand\figureHeightT{3.25}
\newcommand\barWidthT{6}


\vspace{-2mm}
\section{Conclusion}
\vspace{-1mm}
We propose a framework that generates efficient BNN layer-to-device mappings for heterogeneous multiprocessor platforms comprised of CPU and CUDA-capable GPU.
Given a trained BNN model, our proposed \emph{HEP-BNN} framework systematically evaluates the execution time of the model on CPU and on GPU under different parallel configurations.
We evaluate our framework with two BNN architectures on well-known datasets, running on three different types of hardware platforms.
The results show that, across the tested datasets/BNNs and the different hardware platforms, our proposed framework generates mappings for BNN inference which achieve significantly higher speedup compared to a fully-parallelized approach.
Specifically, the efficient parallel configuration from our \emph{HEP-BNN} framework
reduces inference times by up to $2\times$, $2.6\times$ and $11.8\times$, across the tested target hardware respectively.
The generated GPU code from \emph{HEP-BNN} containing the efficient configuration can also then be used for applications using BNN inference in practice.

We believe that our \emph{HEP-BNN} framework will benefit researchers and practitioners to find efficient execution configurations for BNN inference systems using heterogeneous platforms comprised of CPU and GPU.

\vspace{-2mm}
\section*{Acknowledgment}
\vspace{-1mm}
This paper has been supported by Deutsche Forschungsgemeinschaft (DFG) project OneMemory (405422836), by the Collaborative Research Center SFB 876 ``Providing Information by Resource-Constrained Analysis'' (project number 124020371), subproject A1 (http://sfb876.tu-dortmund.de) and by the Federal Ministry of Education and
Research of Germany and the state of NRW as part of
the Lamarr-Institute for ML and AI, LAMARR22B.
This work has received funding by the German Federal Ministry of Education and Research (BMBF) in the course of the 6GEM research hub under grant number 16KISK038.

\vspace{-3mm}
\bibliography{references}
\vspace{-3mm}
\bibliographystyle{ieeetr}

\end{document}

%% file: figures/fi_op_diagram.tex
	\begin{tikzpicture}
	

		\node[minimum height=2cm, minimum width=4cm, draw,fill=white] (Outer2) at (4,7) {};
		\node (SimText) at (4,7.75) {\textbf{Input: BNN model}};
		\node (Sim) at (4,6.9) {
			\includegraphics[scale=0.65]{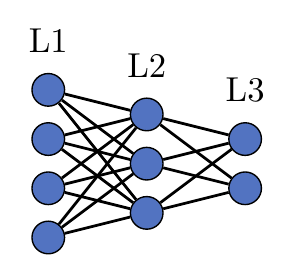}
		};

		\node[minimum height=4cm, minimum width=8cm, draw,fill=white] (Outer3) at (4,3.25) {};
		\node[minimum height=3cm, minimum width=7cm, draw] (Inner1) at (4,3.65) {};
		\node (ResultsText) at (4,4.85) {\textbf{Generate and Infer Configurations}};
		\node (Results) at (4, 3)  {
			\includegraphics[height=3.25cm]{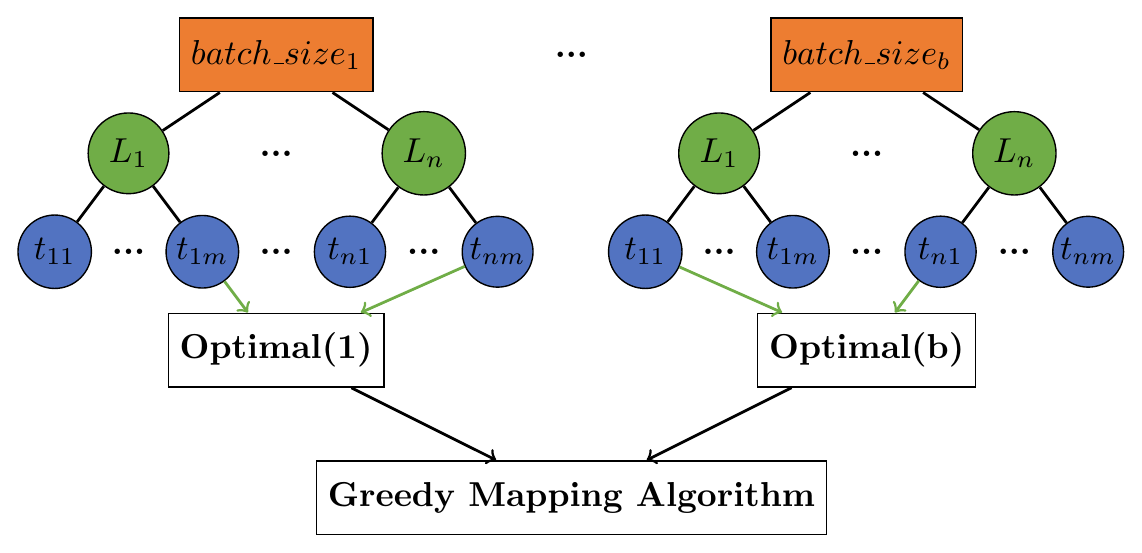}
		};
		
		\node[minimum height = 0.6cm, minimum width=8cm, draw,fill=white] (Outer4) at (4,0.5) {};
		\node (ProbsText) at (4,0.5) {\textbf{Output: Optimal Configurations}};

		\draw[->, thick] (Outer2.south) to (Outer3.north);
		\draw[->, thick] (Outer3.south) to (ProbsText.north);


    \end{tikzpicture}

%% file: figures/tikz/server_cifar.tex
\nextgroupplot[
	    height=\figureHeightT cm,
        width=\figureWidthT cm,
        bar width=1.3*\barWidthT,
        title style={at={(0.5,\titledist)},anchor=north,yshift=-0.1},
        title={CIFAR10 on Server},
        tick pos=left,
        ybar=1pt,
        ymajorgrids = true,
        ylabel = {Latency (s)},
        ylabel near ticks,
        ylabel shift = -0.15cm,
        ytick distance = 150,
        ymax=600,
        symbolic x coords={1, 2, 4, 8, 16, 32, 64, 128},
        scaled y ticks = false,
        enlarge x limits=0.1,
        ymin=0,
        legend cell align=left,
        legend style={
                at={(3.3,-0.5)},
                anchor=south east,
                column sep=1ex
                },
        legend style={
            nodes={scale=0.8, transform shape},
            legend columns=-1,
            },
        legend image post style={scale=0.5}
]

\addplot[draw=myorange, fill=myorange]       coordinates {(1,95.6) (2,91.8) (4,92.5) (8,93.1) (16,93.4) (32,92.7) (64,93.3) (128,93.1)};
\addplot[draw=myred, fill=myred]         coordinates {(1,535.0) (2,282.0) (4,155.0) (8,89.0) (16,128.0) (32,150.0) (64,226.0) (128,269.0)};
\addplot[draw=myblue, fill=myblue]     coordinates {(1,73.5) (2,59.5) (4,53.0) (8,49.0) (16,68.7) (32,96.7) (64,107.0) (128,117.0)};
\addplot[draw=mygreen, fill=mygreen]       coordinates {(1,52.3) (2,45.8) (4,43.5) (8,41.7) (16,41.6) (32,44.0) (64,45.0) (128,45.3)};





%% file: figures/tikz/server_home_cifar.tex
\nextgroupplot[
	    height=\figureHeightT cm,
        width=\figureWidthT cm,
        bar width=1.3*\barWidthT,
        title style={at={(0.5,\titledist)},anchor=north,yshift=-0.1},
        title={CIFAR10 on Laptop},
        tick pos=left,
        ybar=1pt,
        ymajorgrids = true,
        ylabel = {Latency (s)},
        ylabel near ticks,
        ylabel shift = -0.15cm,
        ytick distance=150,
        ymax=700,
        symbolic x coords={1, 2, 4, 8, 16, 32, 64, 128},
        scaled y ticks = false,
        enlarge x limits=0.1,
        ymin=0,
        legend cell align=left,
        legend style={
                at={(3.3,-0.5)},
                anchor=south east,
                column sep=1ex
                },
        legend style={
            nodes={scale=0.8, transform shape},
            legend columns=-1,
            },
        legend image post style={scale=0.5}
]

\addplot[draw=myorange, fill=myorange]       coordinates {(1,104.0) (2,103.0) (4,99.0) (8,100.0) (16,100.0) (32,100.0) (64,100.0) (128,101.0)};
\addplot[draw=myred, fill=myred]         coordinates {(1,671.0) (2,352.0) (4,200.0) (8,202.0) (16,239.0) (32,283.0) (64,314.0) (128,386.0)};
\addplot[draw=myblue, fill=myblue]     coordinates {(1,147.0) (2,112.0) (4,94.0) (8, 144.0) (16, 149.0) (32,151.0) (64,163.0) (128,188.0)};
\addplot[draw=mygreen, fill=mygreen]       coordinates {(1,89.9) (2,76.7) (4,67.1) (8,64.4) (16,60.9) (32,57.6) (64,55.6) (128,55)};





%% file: figures/tikz/server_tx2_cifar.tex
\nextgroupplot[
	    height=\figureHeightT cm,
        width=\figureWidthT cm,
        bar width=1.3*\barWidthT,
        title style={at={(0.5,\titledist)},anchor=north,yshift=-0.1},
        title={CIFAR10 on Jetson TX2},
        tick pos=left,
        ybar=1pt,
        ymajorgrids = true,
        ylabel = {Latency (s)},
        ylabel near ticks,
        ylabel shift = -0.15cm,
        ytick distance=150,
        ymax=800,
        symbolic x coords={1, 2, 4, 8, 16, 32, 64, 128},
        scaled y ticks = false,
        enlarge x limits=0.1,
        ymin=0,
        legend cell align=left,
        legend style={
                at={(3.3,-0.5)},
                anchor=south east,
                column sep=1ex
                },
        legend style={
            nodes={scale=0.8, transform shape},
            legend columns=-1,
            },
        legend image post style={scale=0.5}
]

\addplot[draw=myorange, fill=myorange]       coordinates {(1,469.0) (2,482) (4,481) (8,474) (16,475) (32,475) (64,475) (128,481)};
\addplot[draw=myred, fill=myred]         coordinates {(1,757) (2,708) (4,647) (8,604) (16,572) (32,555) (64,550) (128,552)};
\addplot[draw=myblue, fill=myblue]     coordinates {(1,503) (2,479) (4,447) (8,400) (16,374) (32,365) (64,398) (128,472)};
\addplot[draw=mygreen, fill=mygreen]       coordinates {(1,361) (2,335) (4,321) (8,304) (16,297) (32,301) (64,317) (128,322)};





%% file: figures/tikz/server_fashion.tex
\nextgroupplot[
	    height=\figureHeightT cm,
        width=\figureWidthT cm,
        bar width=1.3*\barWidthT,
        title style={at={(0.5,\titledist)},anchor=north,yshift=-0.1},
        title={Fashion-MNIST on Server},
        tick pos=left,
        ybar=1pt,
        ymajorgrids = true,
        ylabel = {Latency (s)},
        ylabel near ticks,
        ylabel shift = 0.04cm,
        ytick distance=5,
        ymax=25,
        symbolic x coords={1, 2, 4, 8, 16, 32, 64, 128},
        scaled y ticks = false,
        enlarge x limits=0.1,
        ymin=0,
        legend cell align=left,
        legend style={
                at={(3.3,-0.5)},
                anchor=south east,
                column sep=1ex
                },
        legend style={
            nodes={scale=0.8, transform shape},
            legend columns=-1,
            },
        legend image post style={scale=0.5}
]

\addplot[draw=myorange, fill=myorange]       coordinates {(1,2.7) (2,2.5) (4,2.7) (8,3.3) (16,3.2) (32,2.6) (64,2.6) (128,2.7)};
\addplot[draw=myred, fill=myred]         coordinates {(1,23.6) (2,13.3) (4,8.0) (8,5.2) (16,5.4) (32,8.2) (64,8.9) (128,11.3)};
\addplot[draw=myblue, fill=myblue]     coordinates {(1,13.3) (2,8.5) (4,6) (8,4.6) (16,4.2) (32,7.1) (64,8.0) (128,8.8)};
\addplot[draw=mygreen, fill=mygreen]       coordinates {(1,2.6) (2,2.54) (4,2.54) (8, 2.26) (16,2.11) (32,2.12) (64,2.14) (128,2.33)};





%% file: figures/tikz/server_home_fashion.tex

\nextgroupplot[
	    height=\figureHeightT cm,
        width=\figureWidthT cm,
        bar width=1.3*\barWidthT,
        title style={at={(0.5,\titledist)},anchor=north,yshift=-0.1},
        title={Fashion-MNIST on Laptop},
        tick pos=left,
        ybar=1pt,
        ymajorgrids = true,
        ylabel = {Latency (s)},
        ylabel near ticks,
        ylabel shift = 0.04cm,
        ytick distance=25,
        ymax=100,
        symbolic x coords={1, 2, 4, 8, 16, 32, 64, 128},
        scaled y ticks = false,
        enlarge x limits=0.1,
        ymin=0,
        legend cell align=left,
        legend style={
                at={(3.3,-0.5)},
                anchor=south east,
                column sep=1ex
                },
        legend style={
            nodes={scale=0.8, transform shape},
            legend columns=-1,
            },
        legend image post style={scale=0.5}
]

\addplot[draw=myorange, fill=myorange]       coordinates {(1,3.0) (2,2.86) (4,3.2) (8,3.0) (16,3.0) (32,3.5) (64,3.1) (128,3.3)};
\addplot[draw=myred, fill=myred]         coordinates {(1,75.3) (2,46.3) (4,23.8) (8,15.2) (16,18) (32,15.8) (64,15.9) (128,16.9)};
\addplot[draw=myblue, fill=myblue]     coordinates {(1,74.6) (2,33.7) (4,18.1) (8, 11.7) (16, 12.7) (32,11.7) (64,11.8) (128,12.5)};
\addplot[draw=mygreen, fill=mygreen]       coordinates {(1,3.0) (2,2.84) (4,3.1) (8, 3.0) (16, 2.97) (32,3.1) (64,2.87) (128,2.88)};





%% file: figures/tikz/server_tx2_fashion.tex

\nextgroupplot[
	    height=\figureHeightT cm,
        width=\figureWidthT cm,
        bar width=1.3*\barWidthT,
        title style={at={(0.5,\titledist)},anchor=north,yshift=-0.1},
        title={Fashion-MNIST on Jetson TX2},
        tick pos=left,
        ybar=1pt,
        ymajorgrids = true,
        ylabel = {Latency (s)},
        ylabel near ticks,
        ylabel shift = -0.15cm,
        ytick distance=25,
        ymax=150,
        symbolic x coords={1, 2, 4, 8, 16, 32, 64, 128},
        scaled y ticks = false,
        enlarge x limits=0.1,
        ymin=0,
        legend cell align=left,
        legend style={
                at={(0.825,-0.65)},
                anchor=south east,
                column sep=1ex
                },
        legend style={
            nodes={scale=0.8, transform shape},
            legend columns=-1,
            },
        legend image post style={scale=0.5}
]

\addplot[draw=myorange, fill=myorange]       coordinates {(1,11.2) (2,11.5) (4,11.5) (8, 11.7) (16,11.6) (32,11.5) (64,11.5) (128,11.8)};
\addplot[draw=myred, fill=myred]         coordinates {(1,139) (2,87) (4,54.8) (8,37.2) (16,27.7) (32,27.2) (64,25.7) (128,25.6)};
\addplot[draw=myblue, fill=myblue]     coordinates {(1,128) (2,80.8) (4,55.7) (8,38.5) (16,25.9) (32,23.8) (64,24.5) (128,25.4)};
\addplot[draw=mygreen, fill=mygreen]       coordinates {(1,11.1) (2,11.5) (4,11.5) (8,11) (16,9.65) (32,9.58) (64,9.31) (128,9.73)};

\legend{
    CPU only,
    Naive GPU,
    Full parallel GPU,
    Efficient (HEP-BNN Framework),
    }


